\documentclass[11pt]{article}
\usepackage{graphicx,caption2,subfigure}
\makeatletter
\newcommand{\singlespacing}{\let\CS=\@currsize\renewcommand{\baselinestretch}{1.0}\tiny\CS}
\newcommand{\doublespacing}{\let\CS=\@currsize\renewcommand{\baselinestretch}{1.5}\tiny\CS}

\begin{document}
\title {Characteristics of Strange Hadron Production in Some High Energy Collisions and The Role of Power Laws}
\author {S. K. Biswas$^1$\thanks{e-mail:
sunil$\_$biswas2004@yahoo.com},Goutam Sau
$^2$\thanks{e-mail:sau$\_$goutam@yahoo.com}, A.C.Das
Ghosh$^3$\thanks{e-mail: dasghosh@yahoo.co.in}$\&$ S.
Bhattacharyya$^4$\thanks{e-mail: bsubrata@www.isical.ac.in
(Communicating Author).}\\
{\small $^1$ West Kodalia Adarsha Siksha Sadan, New Barrackpore,}\\
 {\small Kolkata-700131, India.}\\
 {\small $^2$ Beramara RamChandrapur High School,}\\
 {\small South 24-Pgs, 743609(WB), India.}\\
{\small $^3$ Department of Microbiology,}\\
{\small Surendranath College, Kolkata-700009, India.}\\
{\small $^4$ Physics and Applied Mathematics Unit$(PAMU)$,}\\
{\small Indian Statistical Institute, Kolkata - 700108, India.}}
\date{}
\maketitle
\bigskip
\bigskip
\singlespacing
\begin{abstract}
Studies on `strange' particle production have always occupied a very
important space in the domain of Particle Physics. This was and is
so, just because of some conjectures about specially abundant or
excess production of `strange' particles, at certain stages and
under certain conditions arising out of what goes by the name of
`Standard' model in Particle Physics. With the help of Hagedornian
power laws we have attempted to understand and interpret here the
nature of the $p_T$-spectra for the strange particle production in a
few high energy nuclear collisions, some interesting
ratio-behaviours and the characteristics of the nuclear modification
factors that are measured in laboratory experiments. After obtaining
and analysing the final results we do not confront any peculiarities
or oddities or extraneous excesses in the properties of the relevant
observables with no left-over problems or puzzles. The model(s) used
by us work(s) quite well for explaining the measured data.
\end{abstract}
\newpage
\doublespacing
\section{Introduction}
 Studying the nature of particle
production in proton-proton collisions is important and interesting
in itself, as it might shed light on the basic mechanism for
production of particles. Besides, it could also serve as a necessary
benchmark for the physics developments in ultra-relativistic heavy
ion collisions\cite{Agaki1,Simone1}. This is specially important at
the large hadron collider (LHC) where the heavy ion programme had
started by November 2010 delivering some preliminary results on some
aspects of strange hadrons produced as the final product in high
energy nuclear collisions and these strange secondaries are supposed
to provide valuable insights into the properties of the
'controversial' system \textit{newly formed}. One of the main
motivation for measuring strange particles in heavy nucleus-nucleus
reactions at LHC is the expectation that their production-rates for
participating nucleon should be enhanced with respect to basic
nucleon-nucleon interactions. Strangeness enhancement has
consistently been proposed as one of the strong diagnostics for a
Quark Gluon Plasma (QGP) state\cite{Koch1,Berndt1}. The enhancement
factor (E) is defined as rapidity-density of multiplicity (yield)
par mean number of nucleon participants [$<N_{part}>$] in heavy ion
collisions, divided by the respective value in p+p collisions. The
requisite information about $<N_{part}>$ etc are to be obtained from
Abelev et al \cite{Abelev1,Abelev2}.
\par~~~~~
\par
 As the strange hadrons are not at all
present in the initial system(A), the question rises very sharply:
how do they make their appearance in the final products. So there
must be some specific reflections on the constituent pictures of the
particles and specifically the nucleons. Besides, enhancement of
strangeness productions was/is one of the powerful diagnostics for
the formation of quark gluon plasma(QGP)in relativistic heavy ion
collisions and the colliders(RHIC).The observations of an increase
of strange baryon production relative to p+p collision in SPS data,
confirmed later at the RHIC studies, has brought excitement in this
area.Besides, the increase of $ p/\pi$ ratio (B) in such collisions
in the non-strange domain had its parallel in the strangeness sector
with the observation of slow rise of the  $\Lambda/k^0$ values.
\par~~~~~
\par
 The organization of this work is as follows.In the section-2 we give an
outline of the model to be applied. In the next section(section-3)we
deliver the results by figures and tables with a short discussion on
the results obtained. In the last chapter, we precisely point out
the conclusions to be arrived at.
\section{The Background In Some Detail And The Working Formulae}
With gradual attainments of larger transverse momenta($p_T$) of the
secondaries in high-$p_T$(hard) interactions, the problems of
deviations from exponential nature of fits on invariant spectra
began to crop up steadily.  Gazdzicki and Gorenstein \cite
{Gazdzicki1} observed rightly that for $p_T>2$ GeV/c, the data
sharply deviate from the exponential nature, for which
Darriulat\cite{Darriulat1} proposed a power law distribution of
certain forms for both $p_T$-spectra and particle multiplicity.
Indeed, for both $p_T$-spectra and multiplicity  such power law
forms have become now the most dominant tools in dealing with the
transverse momentum spectra of all hadrons. Gazdzicki and Gorenstein
showed that the normalised multiplicities and ($m_T ~ \sim ~ p_T$)
spectra of neutral mesons obey the $m_T$-scaling which has had an
approximately power law structure of the form $\sim ~ (m_T)^n$,
where $m_T$ is called transverse mass and is defined by $m_T$=$\surd
p_T^2+m^2$. This scaling behaviour was analogous to that expected in
statistical mechanics, the parameter n plays the role of temperature
and any normalization constant to be used  resembles the system
volume.Thus the basic modification of the statistical approach
needed to reproduce the experimental results on some hadron
production process in $p(\bar p)+p$ interaction in the large $m_T ~
\equiv ~ p_T$ domain is to change the shape of the distribution
functions
 $\exp(-\frac{E^*}{T})$ had to be altered to the power law form as given by
$(\frac{E^*}{\Lambda})^{-n}$ with some changed parameters, viz, a
scale parameter $\Lambda$ and an exponent n, both are assumed to be
common for all hadrons.
\par~~~~~
\par
Let us now dwell, in brief, on the clues to the possible origin of
power laws.  One of the basic features of the hadronuclear
collisions is: irrespective of the initial state,agitations caused
by the impinging
 projectile(be it a parton or particle/nucleon) generate system effects of
 producing avalanches of new kind of partons(called gluons) which form an open
 dissipative system.And these production processes are not at all gradual; rather
 they are very sudden,drastic and complex. And such complex properties and processes
 in nature do generally subscribe to the power-law behaviours.In the recent times,
 it is being propounded consistently that the power law behaviours put into use here are "manifestations
of the dynamics of complex systems whose striking feature is of
showing universal laws characterized by exponents in scale invariant
distributions that happen to be basically independent of the details
in the microscopic dynamics"\cite{S. Lehmann1}. The avalanches
caused by production of excessive number of some new variety of
parton called 'gluons'(the process called 'gluonisation') give rise
to the jettiness of particle production and of cascadisation of the
particle production processes leading to the fractality as is shown
by Sarcevic\cite{Rein1}. These cascades are self-organizing
,self-similar and do just have the fractal behaviour. Driven by the
physical impacts of these well-established factors, in the high
energy collision processes do crop up the several power-laws.And how
such power laws do evolve from exponential origins or roots is
now-a-days being taken care of by the induction of Tsallis entropy
\cite {T.S.Biro1} and a generalisation of Gibbs-Boltzmann statistics
for long-range and multifractal processes.
\par~~~~~
\par
In what follows we are going to choose a specific form of power law
which was previously applied by us in the case of hadron-nucleus
collisions.With a view to accommodating some observed facts for
strange particle production, it is tempting to try to fit the whole
distribution for the inclusive $p_T$-spectra with one single
expression in the form of power law as was done by G. Arnison et
al\cite{Arnison1} and Hagedorn\cite{Hagedorn1}.
\begin{equation}
    E \frac {d^3\sigma}{dp^3}= const.\frac {d(dN/dy)}{2\pi
p_T dp_T}= A (\frac {q} {p_T+q})^n
\end{equation}
where the letters and expressions have their contextual
significance.This parametrization describe the data well over the
entire range of $p_T$.
\par~~~~~
\par
Indeed for $p_T \rightarrow 0, \infty$ we have,
\begin{equation}
(\frac {q}{p_T+q})^n \approx (1-\frac{n}{q} p_T) \approx \left
\{\begin{array}{lll}
exp[\frac{n}{q} p_T] ~~& \mbox{for} & ~ p_T \rightarrow 0 \\
&\mbox{and}&\\
(\frac{q}{p_T})^n ~~ &\mbox{for}&~ p_T \rightarrow \infty
\end{array}
\right.
\end{equation}
 Thus along with impressive fit, which now includes the large $p_T
$ domain, the estimate of $<p_T>$ assumes with the help of
expression(1) :
\begin{equation}
 <p_T>=\frac{\int q/(p_T+q)^n p_T^2 dp_T} {\int q/(p_T+q)^n
p_T dp_T} =\frac {2q} {n-3}
\end{equation}
 So, in clearer terms, let us put the final working formulae here as
follows with substitution of $p_T$ (transverse momentum) as $x$ in
the power-law model\cite{T.Peitzmann1,Albajar1} respectively
\begin{equation}
f(x)=A(1+x/q)^{-n}
\end{equation}
\par There is yet another very important observable called nuclear modification factor (NMF),
denoted here by $R_{CP}$, which for production of any hadron is
defined by \cite{Abelev3}
\begin{equation}\displaystyle{
R_{CP}(p_T)=\frac{[(d^2N/(2\pi dp_Tdp_Tdy))_c/N_{bin}]^{Cenrral}}{[(d^2N/(2\pi dp_Tdp_Tdy))_c/N_{bin}]^{Peripheral}}}
\end{equation}

\doublespacing

\section{Results}
In obtaining the results presented here, no serious statistical
calculational procedure was adopted. The graphs are drawn more as
fitological-cum-phenomenological exercises with mainly statistical
errors in considerations. The experimental data do not provide, in
the most cases, any systematic errors. Data points for the heavy,
high strangeness-valued particles are too scarce; for which the
number of degrees of freedom is too limited for many cases. The
quality of fits to the data indicated in the tables by $\chi^2/ndf$
terms in the columns is attempted to be kept at a modestly
satisfactory value (tending as nearly as possible to unity). And the
figures are drawn by Wgnuplot, wherein there are some inbuilt
statistical procedures and techniques.
\par~~~~~
\par
Quite observably, the results are depicted here in graphical plots.
And the used values of the corresponding parameters for fits are
shown in separate tables.The Fig.1a and Fig.1b show the production
of secondaries $k^0, k^+, k^-$,$\Lambda$ and $\bar{\Lambda}$ in
proton-proton collisions at $\sqrt{s}$=200 GeV at the rapidity
$y<0.5$(Table-1).The figures in Fig.2a and Fig.2b depict the results
for the $\Xi^-, \bar{\Xi}^+ , \Lambda, \bar{\Lambda}$ particle
production for the same collision at the same energy(Table-2).The
cases of $k^+$ and $k^-$ production in gold-gold reaction at the
same energy and at different centralities are reproduced by power
laws in Fig.3a and Fig.3b(Table-3). In Fig.4a and Fig.4b the
production of $\Lambda$ and $\bar{\Lambda}$ are shown in the same
collision
 at the same energy(Table-4). In Fig.5a, Fig.5b and Fig.5c the
 production of cascade, cascade-bar and omega particles production
 are shown at different centralities and at the same energy
 (Table-5). The cases of production of neutral kaons
 and lamda particles in Copper-Copper collision at $\sqrt{s}$=200 GeV are plotted in
 Fig.6a and Fig.6b with reckoning of the parameters
 presented in Table-6 and Table-7.
\par~~~~~
\par
  In Table-8 the values of average transverse momenta ($<p_T>$) for different
 produced secondaries in proton-proton and gold-gold collisions have
 been computed. All these values tally with the similar ranges
 arrived at by experimental measurements. This helps us to obtain for
 us a consistency check-up of the parameter values used for getting
 fits to the data on $<p_T>$-spectra. In Fig.7a and Fig.7b we see,
 the lamda-bar to lamda and cascade-minus to cascade-plus particle
 production cross-section ratio as a function of transverse momentum
 respectively and the ratio gradually fall off with increasing values of $p_T$. In
 Fig.8a and Fig.8b the nuclear modification factors ($R_{CP}$) are
 plotted against transverse momentum for the production of neutral
 meson and lamda particles in copper-copper collision. With the increasing $p_T$, the $R_{CP}$-values
 fall off. In addition, the
 data show the $R_{CP}$ for baryons exhibits a lower fall-off
 compared with that of mesons in intermediate transverse momentum
 region. The experimental data show that the baryon-meson difference
 of $R_{CP}$ disappears at higher $p_T$.
 \par~~~~~
\par
 The data on production of strange particles described here pertain,
in the main, to the 'hard' sector of high energy interactions. And
it is well-known that Hagedornian power law forms which have their
roots in the physics of quantum chromodynamics(QCD) describe hard
particle production in a modestly successful manner for, at least,
the light hadrons of which strange K-mesons constitute a part.But
some of the strange particles have moderately high masses, for which
our objective here was primarily to check whether this generalized
power law form could address the issues of invariant $p_T$-spectra
and some other related observables in a satisfactory manner for all
the strange hadrons. And the outcome is: this study is strongly
affirmative by all indications and yardsticks of actual
performances.
\par~~~~~
\par
The data on strange particles are in general quite sparse. The
errors in measurements are also in most cases quite considerable. If
these limitations are taken into account, the importance of this
comprehensive work, though entirely phenomenological, assumes some
degree of importance. Barring these generalized comments on an
overall basis, we have some specific observations as well, which are
also quite well-merited and are being presented hereafter : (i) The
power indices for all the varieties of strange hadrons lie in the
most cases,in the range of 10-12. This is in accord with the
prescription on the limit set by Brodsky\cite{Brodsky1}, with the
q-values bordering on the values, 2-3. (ii) The $\chi^{2}$/ndf
values for production of cascade and omega particles have suffered
quantitatively due to very small values of the number of degrees of
freedom. (iii) Thirdly, the q and n values do not exhibit any clear
centrality-dependence or the mass-dependence of the observed heavy
secondaries. (iv) However, we cannot ascertain at this moment the
properties of these parameters with regard to the nature of their
energy-dependence(s), if any. (v) The average momentum values of
these measured heavy strange baryons are found to be quantitatively
compatible with other non-strange light hadrons, though the
expression for the average transverse momentum is not very
rigorously derived, for which reliability of expression (3) is
certainly limited. (vi) Some ratio-values shown by Fig.7a and Fig.7b
are modestly well-described. (vii) However, the values of the
nuclear modification factor, denoted by $R_{CP}$, are not reproduced
satisfactorily, especially on the lower-side of the $p_T$-values.
But this is no wonder, as the used power laws are suited to
high-$p_T$ values as was remarked above very concretely. (viii)
Still, with one of the simplest approaches, that we have succeeded
in explaining the characteristics of a large bulk of data on these
rare hadrons is certainly quite stimulating to and encouraging for
us.
\section{Concluding Remarks}
Let us now sum up;
\par~~~~~
\par (i) The used power laws explain quite well
the measured data on the observables like, $p_T$-spectra, some
ratio-behaviours and the nuclear modification factors; so none of
the questions related to suppression or enhancement is
consequential. (ii) The centrality- dependences of the $p_T$-spectra
of strange hadrons too are well-reproduced. (iii) The essential
physical content of the power-law models is the modest observance of
the $p_T$-scaling (as is reflected in the $\sim$ $p_T$/$p_0$ term).
And in terms of the functional efficacy, this model seems, so far,
to be at par, if not better, with all other numerous existing
approaches within the frameworks of either the Hadronic transport
models or the statistical models\cite{Blume1}-\cite{Torrieri2}. (iv)
The model obviously bears no relationship with the concept of the
quark-gluon plasma (QGP), which is, still, just a conjecture with so
far no clear and concrete experimental support. (v) In a conclusive
vein, this has to be asserted that we observe nothing too strange
about `strangeness' production in high energy interactions.
\par~~~~~
\par
The latter two points in the preceding paragraph highlight, in the
main, the novelty of this study from a global viewpoint against the
background of the current trends and streaming ideas in the present
day Particle Physics.
\par~~~~~
\par\textbf{Acknowledgement}
\par The authors express their thankful gratitude to the honourable
referee for making some valuable comments and suggestions.

\newpage
\singlespacing

\newpage
{\singlespacing{
\begin{table}
\begin{center}
\begin{small}
\caption{ Numerical values of the fit parameters of power law
equation for  keon and lamda production in p-p collisions at
$\surd{s_{NN}}$=200 GeV, $p_T$ = 0 to 5GeV/c}
\begin{tabular}{|c|c|c|c|c|}\hline
Sesondaries  & $A$ & $q$ & $n$ &$\frac{\chi^2}{ndf}$\\
\hline $K^{0}_{s}$  & $0.563 \pm 0.023$ &
$3.108 \pm
0.025$ & $15.005 \pm 0.032$ & $20.707/17$\\
\hline $K^+$  & $1.067 \pm 0.032$ &
$1.581 \pm
0.055$ & $10.000 \pm 0.209$ & $10.943/9$ \\
\hline $K^-$ & $0.066 \pm 0.008$ &
$2.895 \pm
0.026$ & $15.116 \pm 0.405$ & $0.584/6$ \\
\hline $\Lambda$  & $0.273 \pm 0.036$ &
$3.092 \pm
0.068$ & $15.007 \pm 0.074$ & $18.393/15$ \\
\hline ${\Lambda}bar$  & $0.029 \pm 0.001$ &
$3.010 \pm
0.068$ & $15.016 \pm 0.068$ & $15.673/15$ \\
\hline
\end{tabular}
\end{small}
\end{center}
\end{table}
\begin{table}
\begin{center}
\begin{small}
\caption{ Numerical values of the fit parameters of power law
equation for lamda and cascade particle production in p-p collisions
at $\surd{s_{NN}}$=200 GeV, $p_T$ = 0 to 5GeV/c}
\begin{tabular}{|c|c|c|c|c|}\hline
Sesondaries  & $A$ & $q$ & $n$ &$\frac{\chi^2}{ndf}$\\
\hline
$\Lambda$  & $1.848 \pm 0.094$ &
$1.306 \pm
0.061$ & $10.417 \pm 0.244$ & $26.883/12$ \\
\hline $\Lambda{bar}$  & $0.541 \pm 0.026$ &
$1.910 \pm
0.068$ & $12.005 \pm 0.061$ & $36.675/13$ \\
\hline
$\Xi^-$  & $0.022 \pm 0.001$ &
$1.804 \pm
0.023$ & $10.033 \pm 0.119$ & $21.4/8$ \\
\hline ${\Xi^+}bar$  & $0.063 \pm 0.002$ &
$2.301 \pm
0.015$ & $9.951 \pm 0.466$ & $24.246/9$ \\
\hline
\end{tabular}
\end{small}
\end{center}
\end{table}
\begin{table}
\begin{center}
\begin{small}
\caption{ Numerical values of the fit parameters of Power Law
equation for keon production($k^+,k^-$) in Au-Au collisions at
$\surd s_{NN}$=200GeV at different Centrality, $p_T$ = 0 to 2 GeV/c}
\begin{tabular}{|c|c|c|c|c|c|}\hline
 Secondaries&Centrality(\%) & $A$ & $q$ & $n$ & $\frac{\chi^2}{ndf}$\\
\hline$k^+$&    0-5$\%$ & $120.114 \pm 8.341$ & $2.508 \pm
0.057$ & $10.376 \pm 0.172$ & $4.555/14$ \\
 \hline  &$ 20-30\%$ & $47.371 \pm 2.28$ & $1.998
\pm
0.029$ & $9.989 \pm 0.112$ & $2.477/15$ \\
 \hline  &$ 40-50\%$ & $16.727 \pm 2.703$ & $1.843
\pm
0.082$ & $10.006 \pm 0.111$ & $1.294/12$ \\
 \hline
&$60-70\% $ & $3468.89 \pm 253.7$ & $2.030 \pm
0.033$ & $11.981 \pm 0.154$ & $3.022/13$ \\
\hline  $ k^-$& 0-5$\%$ & $435.675 \pm 15.85$ & $2.286 \pm
0.076$ & $12.00 \pm 0.264$ & $1.032/15$ \\
 \hline  &$ 20-30\%$ & $68.462 \pm 6.505$ & $2.379
\pm
0.061$ & $12.003 \pm 0.07$ & $0.246/12$ \\
 \hline  &$ 40-50\%$ & $25.454 \pm 0.901$ & $2.239
\pm
0.084$ & $12.052 \pm 0.297$ & $1.566/14$ \\
 \hline
 &$60-70\%$&$5.058\pm0.283$&$2.016\pm0.101$&$11.819\pm0.382$&$3.356/14$\\\hline
 \end{tabular}
\end{small}
\end{center}
\end{table}
\begin{table}
\begin{center}
\begin{small}
\caption{ Numerical values of the fit parameters in Power Law form
for ($\Lambda$ and $\bar{\Lambda}$) production
 in Au-Au collisions at $\surd s_{NN}$=200GeV at
 $p_T$ = 1 to 5 GeV/c
ranges and for various Centrality-values as given below.}
\begin{tabular}{|c|c|c|c|c|c|}\hline
 Secondaries&Centrality(\%) & $A$ & $q$ & $n$ & $\frac{\chi^2}{ndf}$\\
\hline$\Lambda$&    0-5$\%$ & $63.570 \pm 4.841$ & $3.542 \pm
0.105$ & $16.124 \pm 0.140$ & $10.064/18$ \\
 \hline  &$ 10-20\%$ & $772.323 \pm 48.46$ & $2.000
\pm
0.087$ & $13.870 \pm 0.259$ & $5.778/14$ \\
 \hline  &$ 20-40\%$ & $141.42 \pm 11.33$ & $2.002
\pm
0.023$ & $12.815 \pm 0.099$ & $7.512/12$ \\
 \hline
&$40-60\% $ & $34.334 \pm 2.317$ & $2.025 \pm
0.109$ & $12.820 \pm 0.308$ & $7.238/12$ \\
\hline &$60-80\% $ & $19.811 \pm 0.968$ & $2.000 \pm
0.014$ & $11.869 \pm 0.235$ & $14.913/12$ \\
\hline ${\Lambda}bar $& 0-5$\%$ & $198.233 \pm 0.269$ & $2.000 \pm
0.014$ & $10.765 \pm 0.269$ & $7.588/10$ \\
 \hline  &$ 10-20\%$ & $29.161 \pm 1.727$ & $2.002
\pm
0.121$ & $12.003 \pm 0.072$ & $17.298/10$ \\
 \hline  &$ 20-40\%$ & $122.194 \pm 4.406$ & $2.899
\pm
0.014$ & $15.001 \pm 0.054$ & $3.589/8$ \\
 \hline
 &$40-60\%$&$25.183\pm0.646$&$2.500\pm0.008$&$12.888\pm0.028$&$9.902/12$\\\hline
 &$60-80\%$&$10.994\pm0.788$&$2.001\pm0.022$&$11.995\pm0.088$&$31.900/13$\\\hline
 \end{tabular}
\end{small}
\end{center}
\end{table}
\begin{table}
\begin{center}
\begin{small}
\caption{ Numerical values of the fit parameters in Power Law
equation for cascade-minus($\Xi^-$), cascade-plus bar($\Xi^{+}bar$)
and $\Omega^{-}+\Omega^{+}$ particle production in Au-Au collisions
at $\surd s_{NN}$=200GeV at different $p_T$-values = 1 to 5 GeV/c
for various Centrality values.}
\begin{tabular}{|c|c|c|c|c|c|}\hline
 Secondaries&Centrality(\%) & $A$ & $q$ & $n$ & $\frac{\chi^2}{ndf}$\\
\hline$\Xi^-$&    0-5$\%$ & $213.632 \pm 16.06$ & $1.344 \pm
0.015$ & $10.042 \pm 0.271$ & $14.593/8$ \\
 \hline  &$ 10-20\%$ & $1019.24 \pm 81.73$ & $1.001
\pm
0.010$ & $10.017 \pm 0.058$ & $20.800/6$ \\
 \hline  &$ 20-40\%$ & $2978.87 \pm 155.4$ & $1.000
\pm
0.006$ & $11.003 \pm 0.001$ & $6.138/6$ \\
 \hline
&$40-60\% $ & $16.505 \pm 0.288$ & $1.500 \pm
0.003$ & $10.019 \pm 0.055$ & $0.675/7$ \\
\hline &$60-80\% $ & $3.065 \pm 0.339$ & $1.503 \pm
0.027$ & $10.001 \pm 0.111$ & $16.968/5$ \\
\hline $\Xi^{+}bar $& 0-5$\%$ & $1531.25 \pm 144$ & $1.001 \pm
0.012$ & $9.996 \pm 0.070$ & $18.845/7$ \\
 \hline  &$ 10-20\%$ & $5248.25 \pm 468.3$ & $0.801
\pm
0.008$ & $10.011 \pm 0.053$ & $22.498/6$ \\
 \hline  &$ 20-40\%$ & $735.176 \pm 31.68$ & $0.999
\pm
0.005$ & $9.996 \pm 0.031$ & $24.258/8$ \\
 \hline
 &$40-60\%$&$233.981\pm15.39$&$0.999\pm0.008$&$9.997\pm0.047$&$16.173/7$\\\hline
 &$60-80\%$&$2.993\pm0.188$&$1.496\pm0.014$&$10.006\pm0.057$&$4.049/4$\\\hline
 $\Omega^{-}+\Omega^{+}$&$0-5\%$&$1.248\pm0.094$&$3.018\pm0.055$&$9.988\pm0.139$&$2.073/3$\\\hline
 &$20-40\%$&$2.970\pm0.229$&$1.997\pm0.028$&$9.978\pm0.100$&$1.304/2$\\\hline
 &$40-60\%$&$0.342\pm0.037$&$2.001\pm0.043$&$9.235\pm0.131$&$4.755/3$\\\hline
 \end{tabular}
\end{small}
\end{center}
\end{table}
\begin{table}
\begin{center}
\begin{small}
\caption{ Numerical values of the fit parameters of Power Law
equation for keon production($k^{0}_{s}$) in Cu-Cu collisions at
$\surd s_{NN}$=200GeV at different $p_T$-values = 1 to 9 GeV/c} and
for several Centrality domains.
\begin{tabular}{|c|c|c|c|c|c|}\hline
 Secondaries&Centrality(\%) & $A$ & $q$ & $n$ & $\frac{\chi^2}{ndf}$\\
\hline$k^{0}_{s}$&    0-10$\%$ & $112.079 \pm 4.169$ & $2.020 \pm
0.035$ & $12.229 \pm 0.094$ & $6.675/21$ \\
 \hline  &$ 10-20\%$ & $67.284 \pm 2.154$ & $1.921
\pm
0.031$ & $11.712 \pm 0.477$ & $6.068/21$ \\
 \hline  &$ 20-30\%$ & $24.860 \pm 0.568$ & $2.879
\pm
0.030$ & $14.155 \pm 0.262$ & $3.667/21$ \\
 \hline
&$30-40\% $ & $17.099 \pm 0.286$ & $2.565 \pm
0.070$ & $13.181 \pm 0.0.245$ & $6.230/21$ \\
\hline  & 40-50$\%$ & $18.504 \pm 0.598$ & $1.985 \pm
0.036$ & $11.583 \pm 0.476$ & $7.738/21$ \\
 \hline  &$ 50-60\%$ & $12.159 \pm 0.536$ & $2.000
\pm
0.006$ & $11.551 \pm 0.054$ & $7.842/21$ \\
 \hline
\end{tabular}
\end{small}
\end{center}
\end{table}
\begin{table}
\begin{center}
\begin{small}
\caption{ Numerical values of the fit parameters of Power Law
equation for lamda particle($\Lambda$) production in Cu-Cu
collisions at $\surd s_{NN}$=200GeV at different Centrality, $p_T$ =
1 to 7 GeV/c}
\begin{tabular}{|c|c|c|c|c|c|}\hline
 Secondaries&Centrality(\%) & $A$ & $q$ & $n$ & $\frac{\chi^2}{ndf}$\\
\hline$\Lambda$&    0-10$\%$ & $1091.23 \pm 47.1$ & $2.000 \pm
0.009$ & $14.428 \pm 0.166$ & $2.635/12$ \\
 \hline  &$ 10-20\%$ & $139.73 \pm 12.91$ & $2.034
\pm
0.103$ & $13.033 \pm 0.275$ & $15.067/14$ \\
 \hline  &$ 20-30\%$ & $105.009 \pm 5.974$ & $2.024
\pm
0.073$ & $12.996 \pm 0.193$ & $16.425/17$ \\
 \hline
&$30-40\% $ & $46.450 \pm 1.523$ & $2.026 \pm
0.081$ & $12.629 \pm 0.216$ & $24.731/17$ \\
\hline  &40-50$\%$ & $27.007 \pm 1.798$ & $2.001 \pm
0.018$ & $12.340 \pm 0.215$ & $16.803/15$ \\
 \hline  &$ 50-60\%$ & $24.819 \pm 2.077$ & $2.029
\pm
0.023$ & $13.042 \pm 0.097$ & $17.437/14$ \\
 \hline
\end{tabular}
\end{small}
\end{center}
\end{table}
\begin{table}
\begin{center}
\caption{Calculated values of average transverse momentum $<p_T>$
for $y_{cm}<0.5$  at $\surd s_{NN}$=200GeV .}
\begin{tabular}{|c|c|c|c|}\hline
Nature of Collisions & Secondaries & Value of
$<p_T>$ in GeV/c \\ \hline p-p Collisions &  $k^{0}_{s} $& 0.52\\
& $k^{+}$ & 0.45\\  &
 $k^{-}$& 0.48\\ & $\Lambda$& 0.52\\
 & $\bar{\Lambda}$ & 0.50\\
  & $\Xi^-$&0.51\\& $\Xi^{+}bar$&0.86\\
 \hline
Au-Au Collisions & $k^{+}$ &0.68\\
(Central)&$k^{-}$&0.50\\&$\Lambda$
&0.54\\&$\bar{\Lambda}$
&0.52\\&$\Xi^-$&0.38\\
&$\Xi^{+}bar$ &0.28\\&$\Omega^{-}+\Omega^{+}$ &0.86\\\hline
\end{tabular}
\end{center}
\end{table}

\newpage
\begin{figure}
\subfigure[]{
\begin{minipage}{.5\textwidth}
\centering
\includegraphics[width=2.5in]{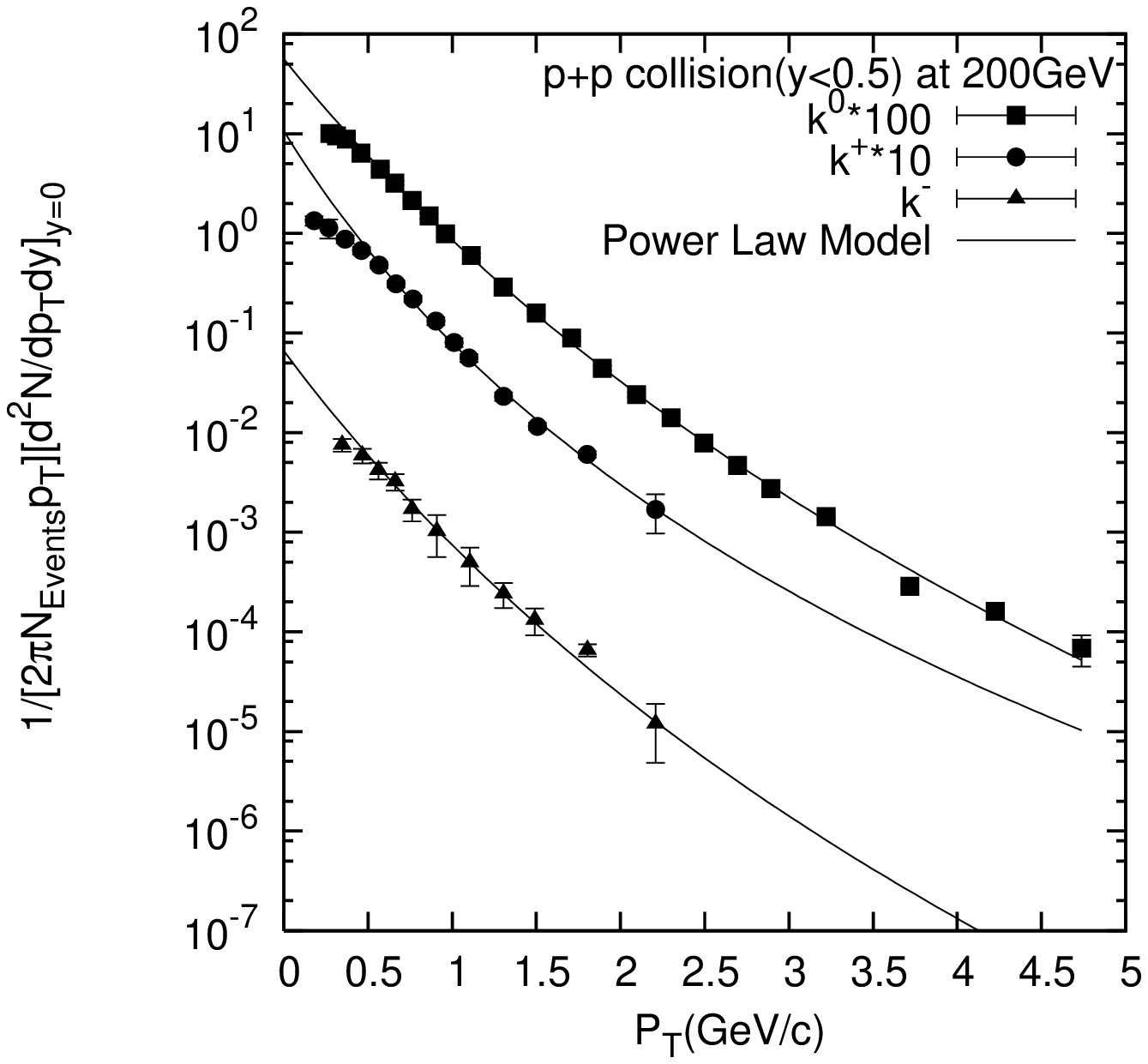}
\end{minipage}}%
\subfigure[]{
\begin{minipage}{.5\textwidth}
\centering
\includegraphics[width=2.5in]{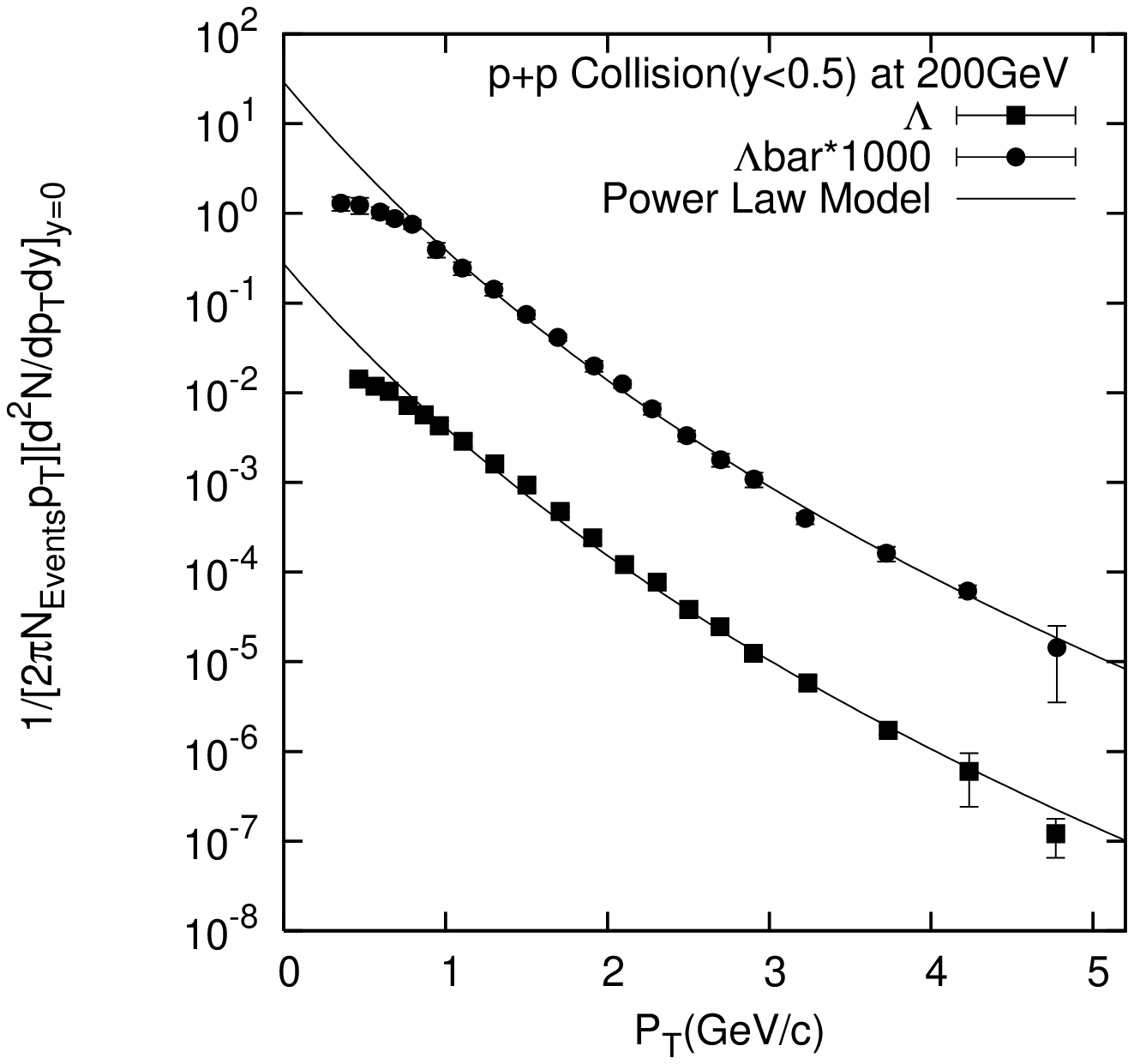}
\end{minipage}}%
\caption{Transverse momentum spectra for production of
keon ($K^0, K^+, K^-$), lamda($\Lambda$), lamdabar($\bar{\Lambda}$),
 cascade minus($\Xi^-$), cascade plus bar(${\Xi^+}bar$) particles in
pp collisions at $\surd{s}$ = 200GeV. The experimental data are taken from Ref.\cite{Abelev1}.
The solid curves are fits for power-law model.}
\end{figure}
\begin{figure}
\subfigure[]{
\begin{minipage}{.5\textwidth}
\centering
\includegraphics[width=2.5in]{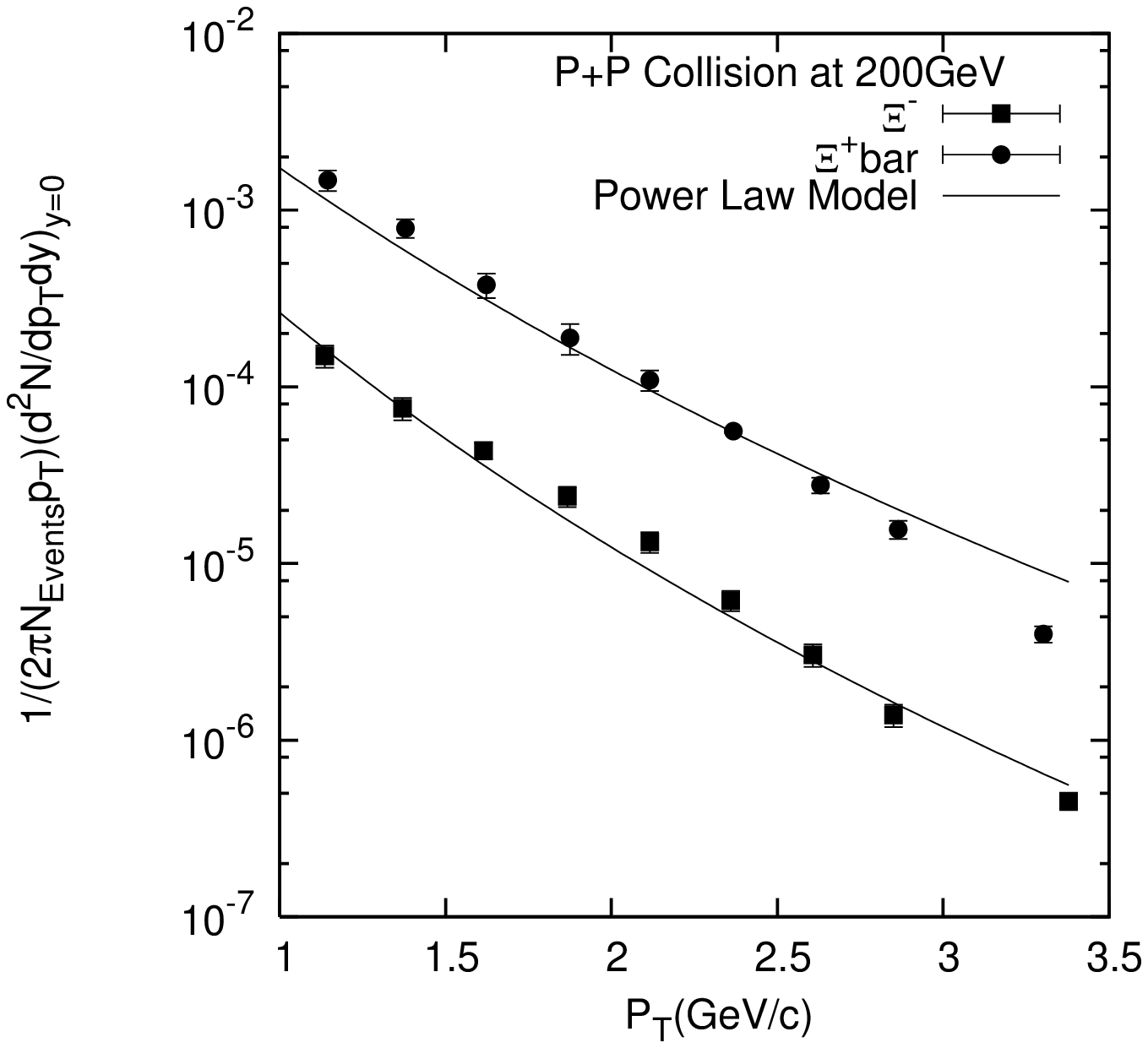}
\end{minipage}}%
\subfigure[]{
\begin{minipage}{.5\textwidth}
\centering
 \includegraphics[width=2.5in]{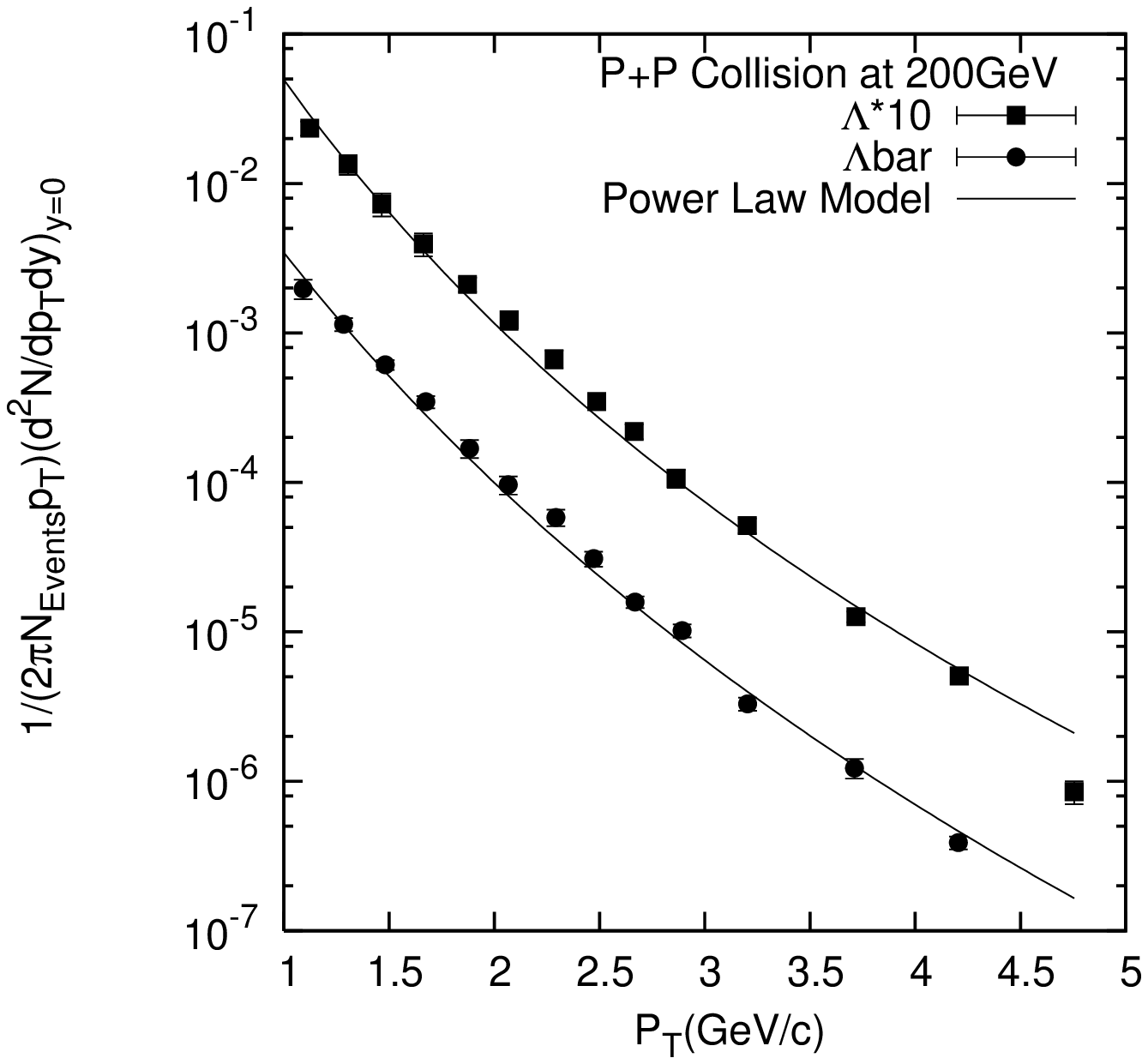}
 \end{minipage}}%
\caption{Transverse momentum spectra for production
of cascade($\Xi^-,\Xi^+bar$), lamda and lamdabar particles in
pp collisions at $\surd{s}$=200GeV. The experimental data are taken from Ref.\cite{Heinz1}.
The solid curves are fits for power-law model.}
\end{figure}
\begin{figure}
\subfigure[]{
\begin{minipage}{.5\textwidth}
  \centering
\includegraphics[width=2.5in]{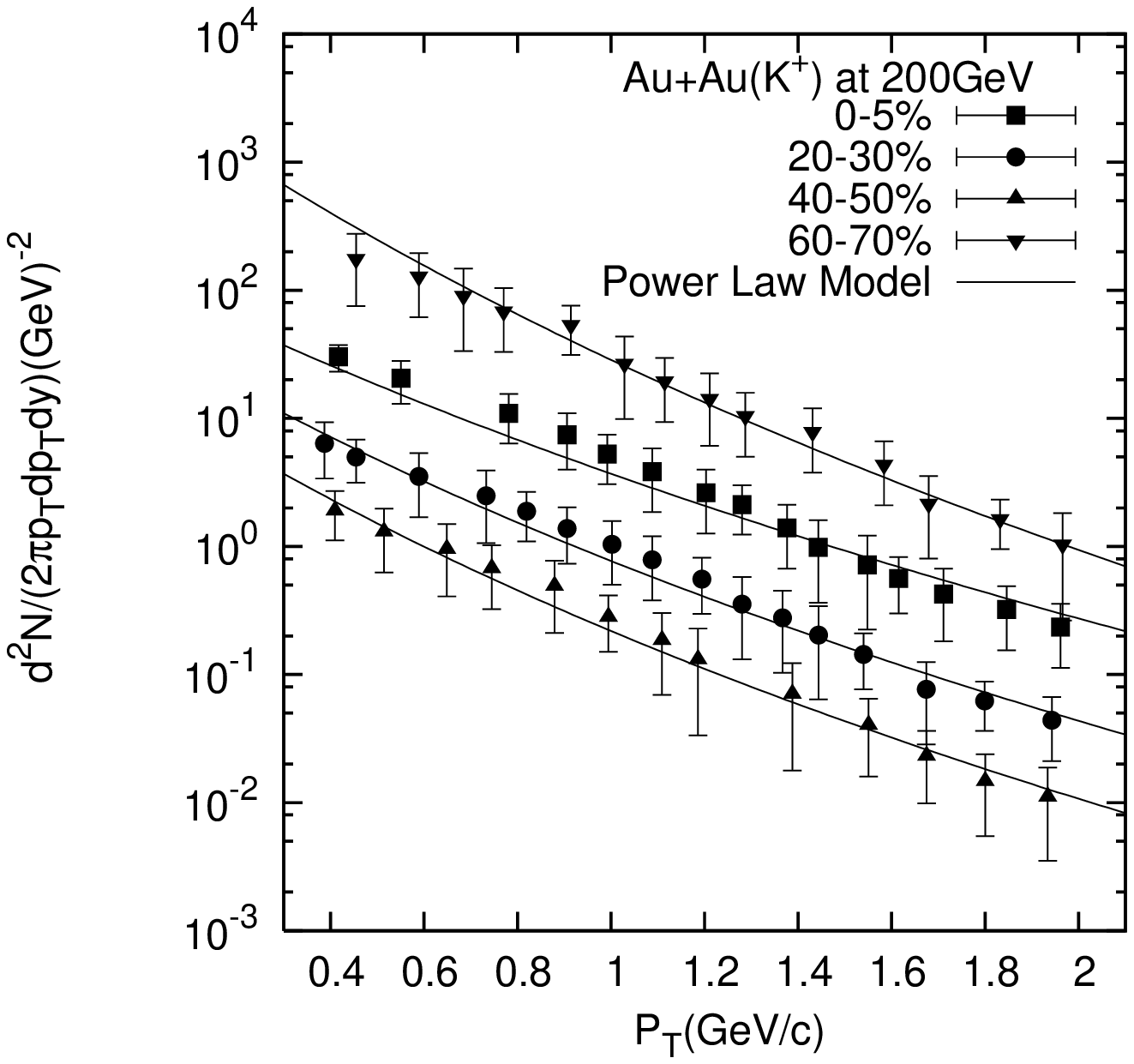}
\end{minipage}}%
\subfigure[]{
\begin{minipage}{.5\textwidth}
\centering
\includegraphics[width=2.5in]{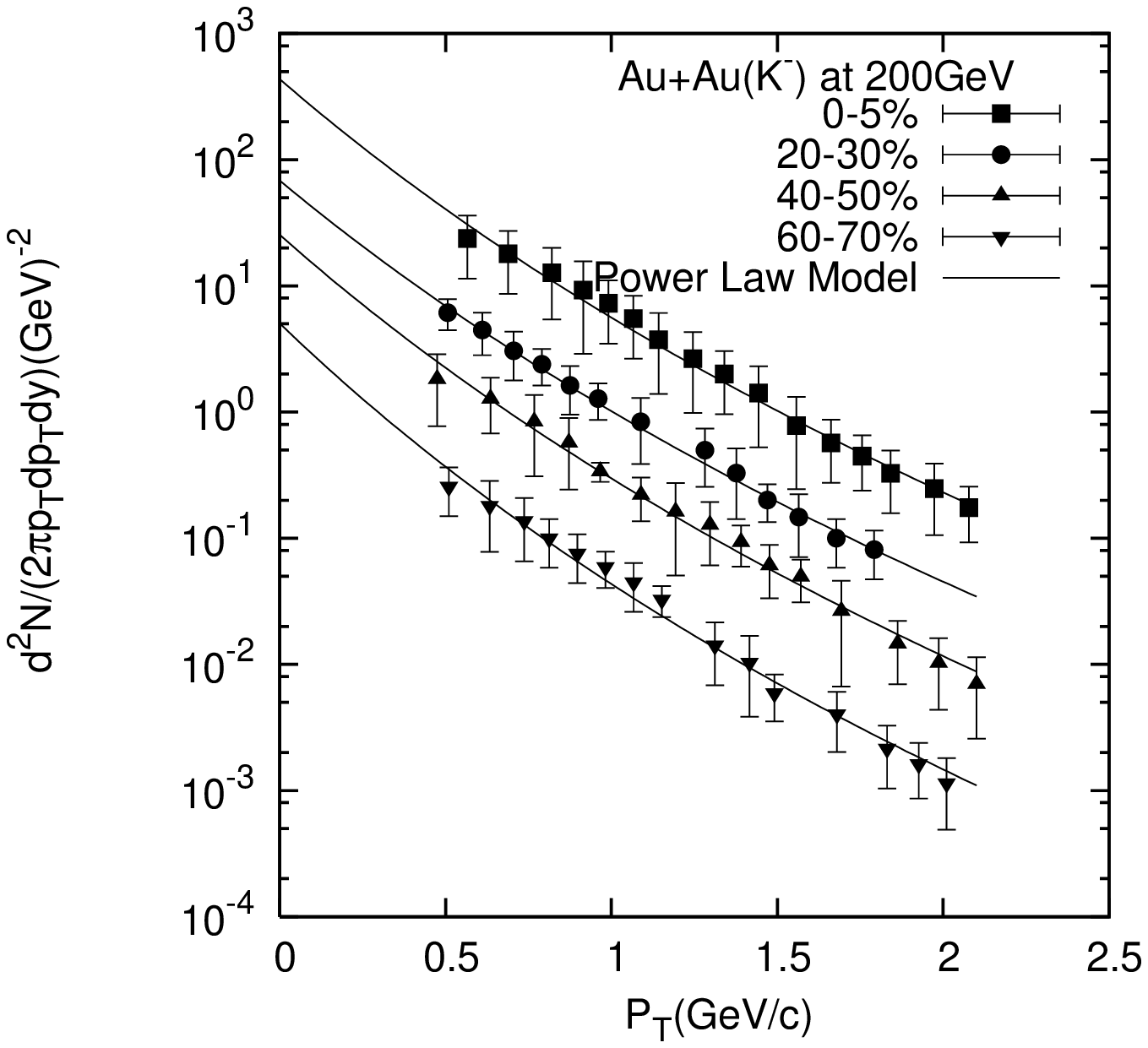}
\end{minipage}}%
\caption{Transverse momentum spectra for production of kaon($K^+,K^-$)at different
centrality at $\surd{s}$=200GeV in Au-Au collisions. The experimental data are taken from Ref.\cite{Wang1} The
solid curves are fits for power-law model.}
\end{figure}
\begin{figure}
\subfigure[]{
\begin{minipage}{.5\textwidth}
  \centering
\includegraphics[width=2.5in]{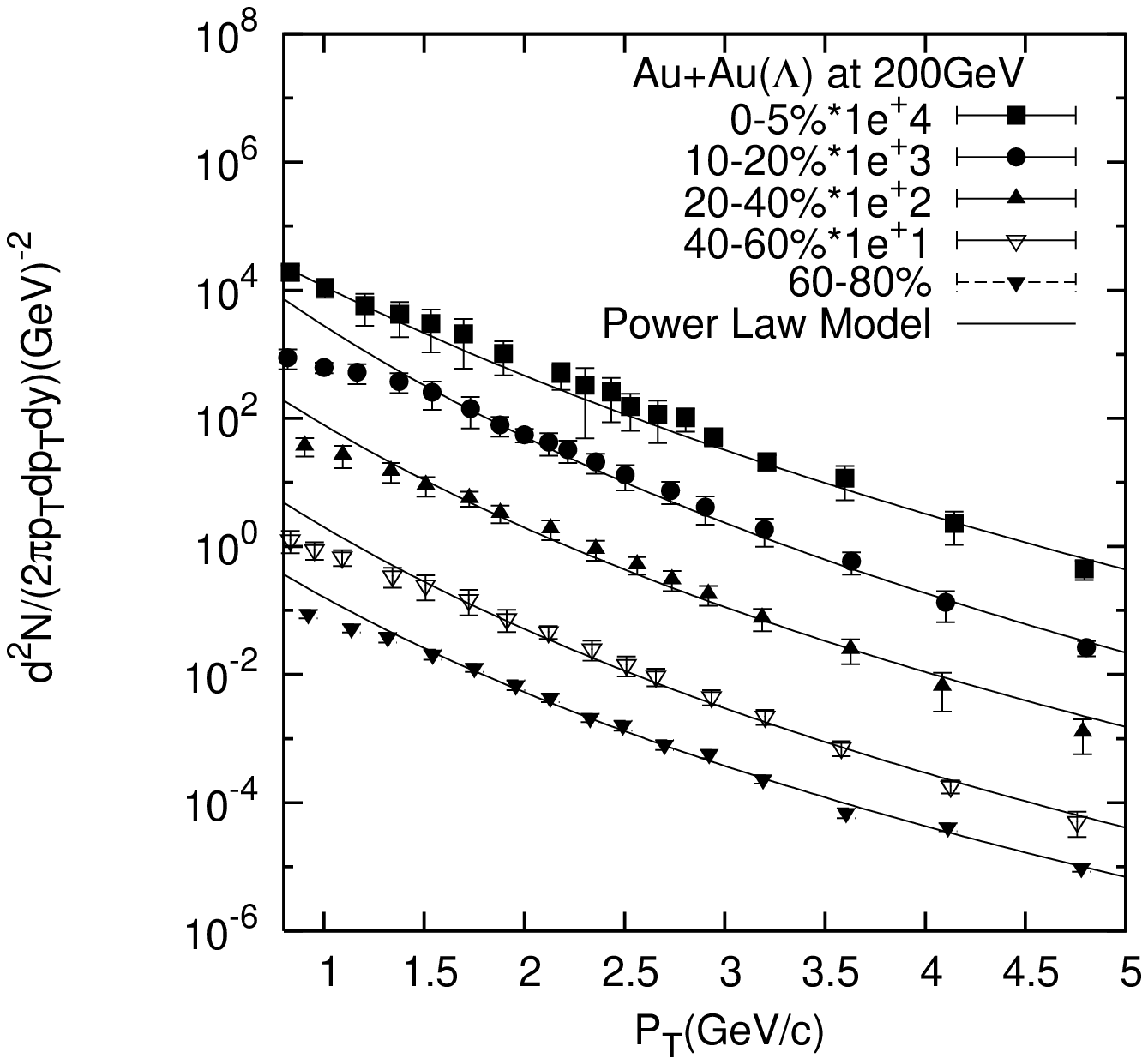}
\end{minipage}}%
\subfigure[]{
\begin{minipage}{.5\textwidth}
\centering
\includegraphics[width=2.5in]{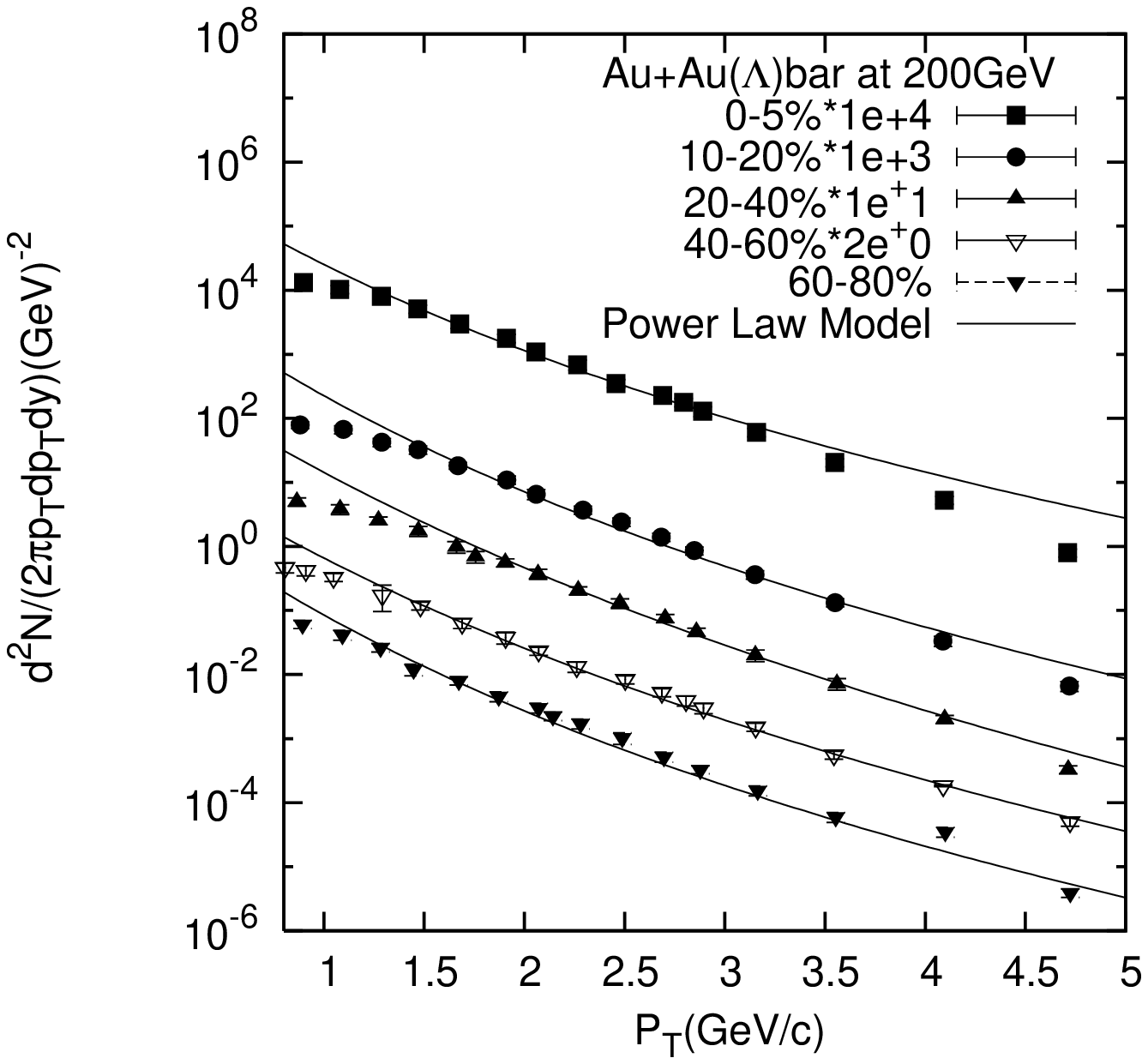}
\end{minipage}}%
\caption{Transverse momentum spectra for production of lamda($\Lambda$) and lamda bar($\bar{\Lambda}$) particles
at different centrality in Au-Au collisions. The experimental data are taken from Ref.\cite{Wang1} The
solid curves are fits for power-law model.}
\end{figure}
\begin{figure}
\subfigure[]{
\begin{minipage}{.5\textwidth}
\centering
\includegraphics[width=2.5in]{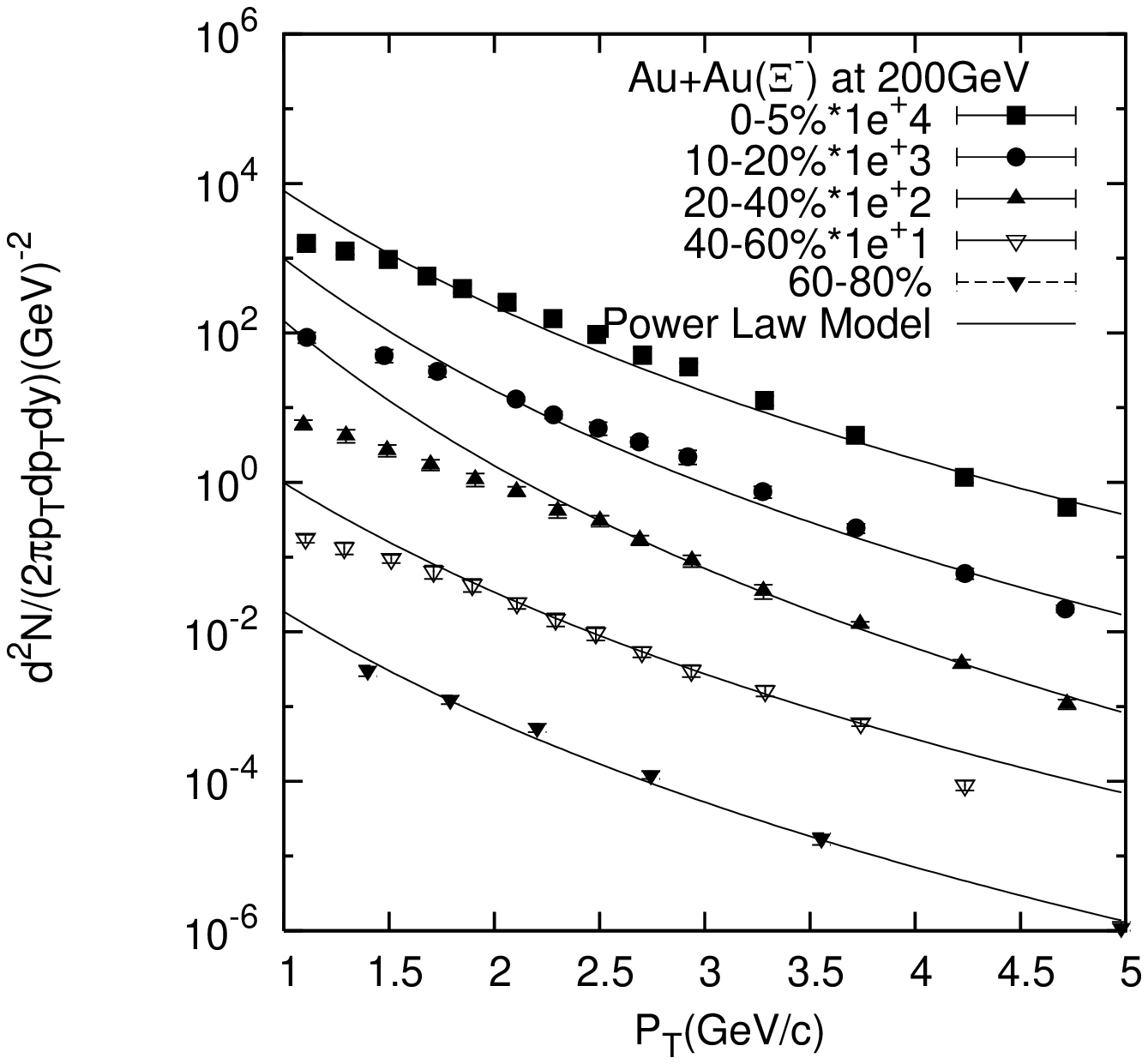}
\end{minipage}}%
\subfigure[]{
\begin{minipage}{.5\textwidth}
\centering
 \includegraphics[width=2.5in]{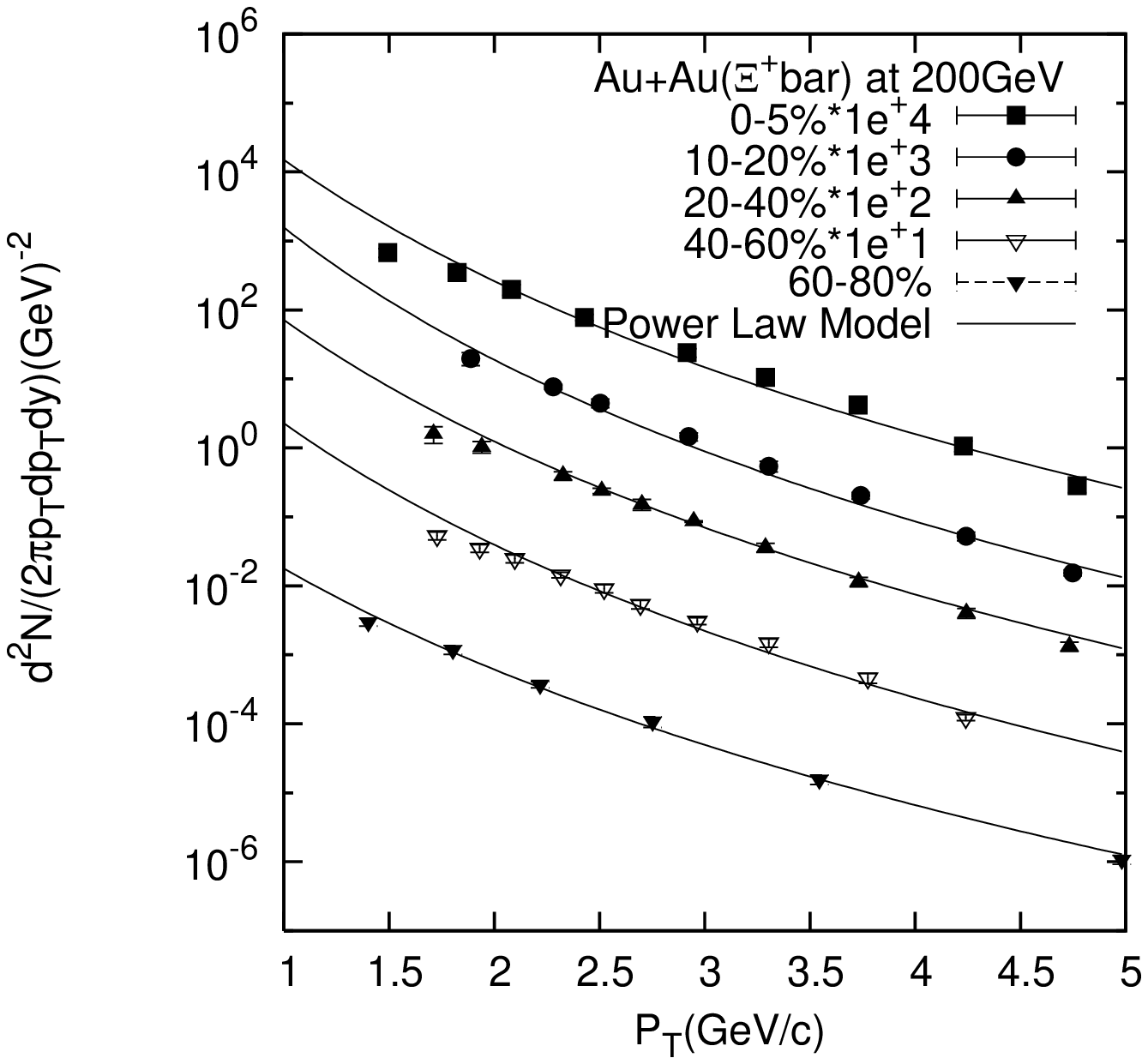}
 \end{minipage}}%
\vspace{0.01in} \subfigure[]{
\begin{minipage}{1\textwidth}
  \centering
\includegraphics[width=2.5in]{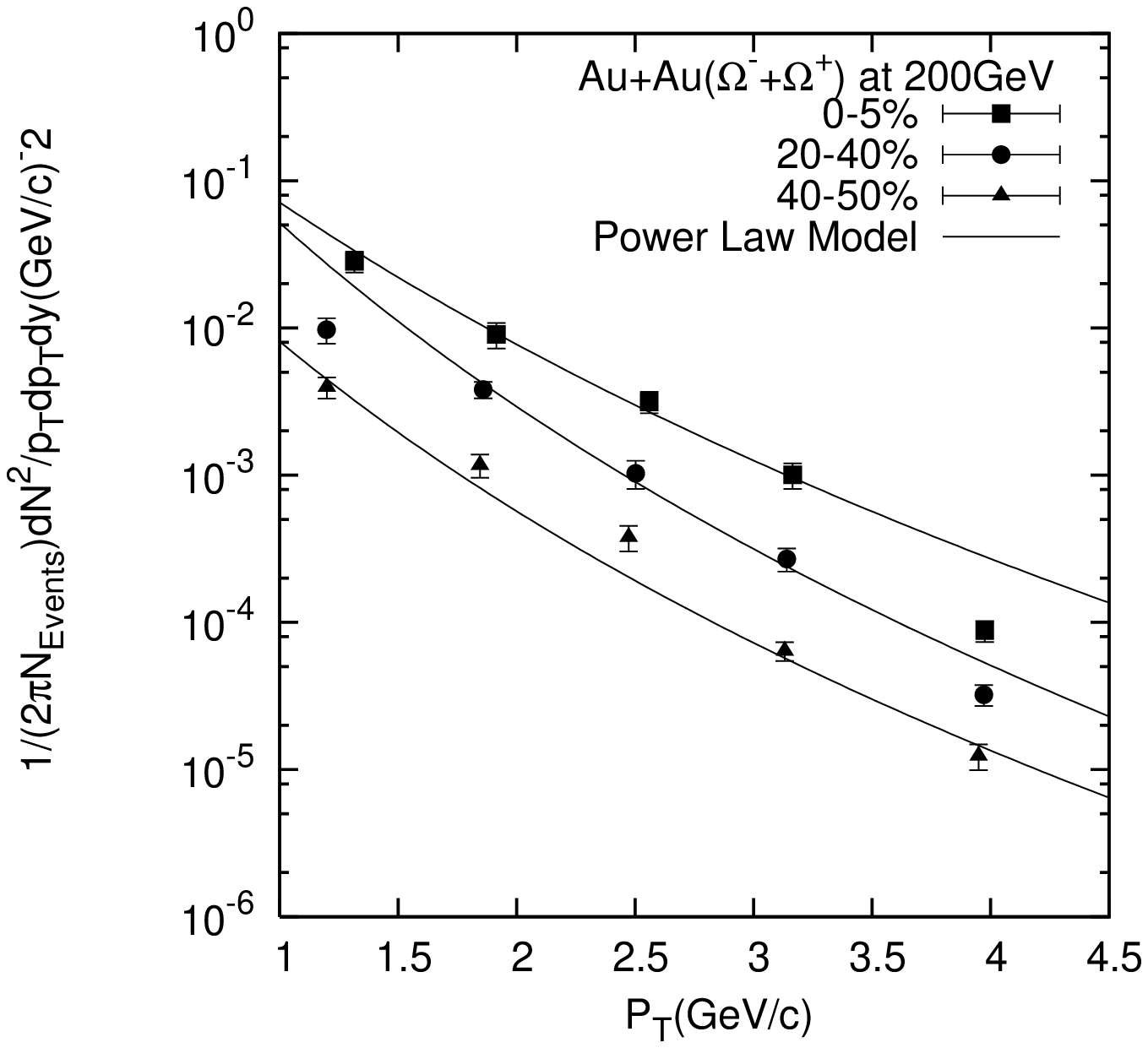}
\end{minipage}}%
\caption{Transverse momentum spectra for production of cascade minus($\Xi^-$),cascade plus bar($\Xi^+bar$)
and omega particles($\Omega^-+\Omega^+$) at different centrality in Au-Au collisions at $\surd{s}$=200GeV .
The experimental data are taken from Ref.\cite{Adams1} The solid curves are fits for power-law model.}
\end{figure}
\begin{figure}
\subfigure[]{
\begin{minipage}{.5\textwidth}
  \centering
\includegraphics[width=2.5in]{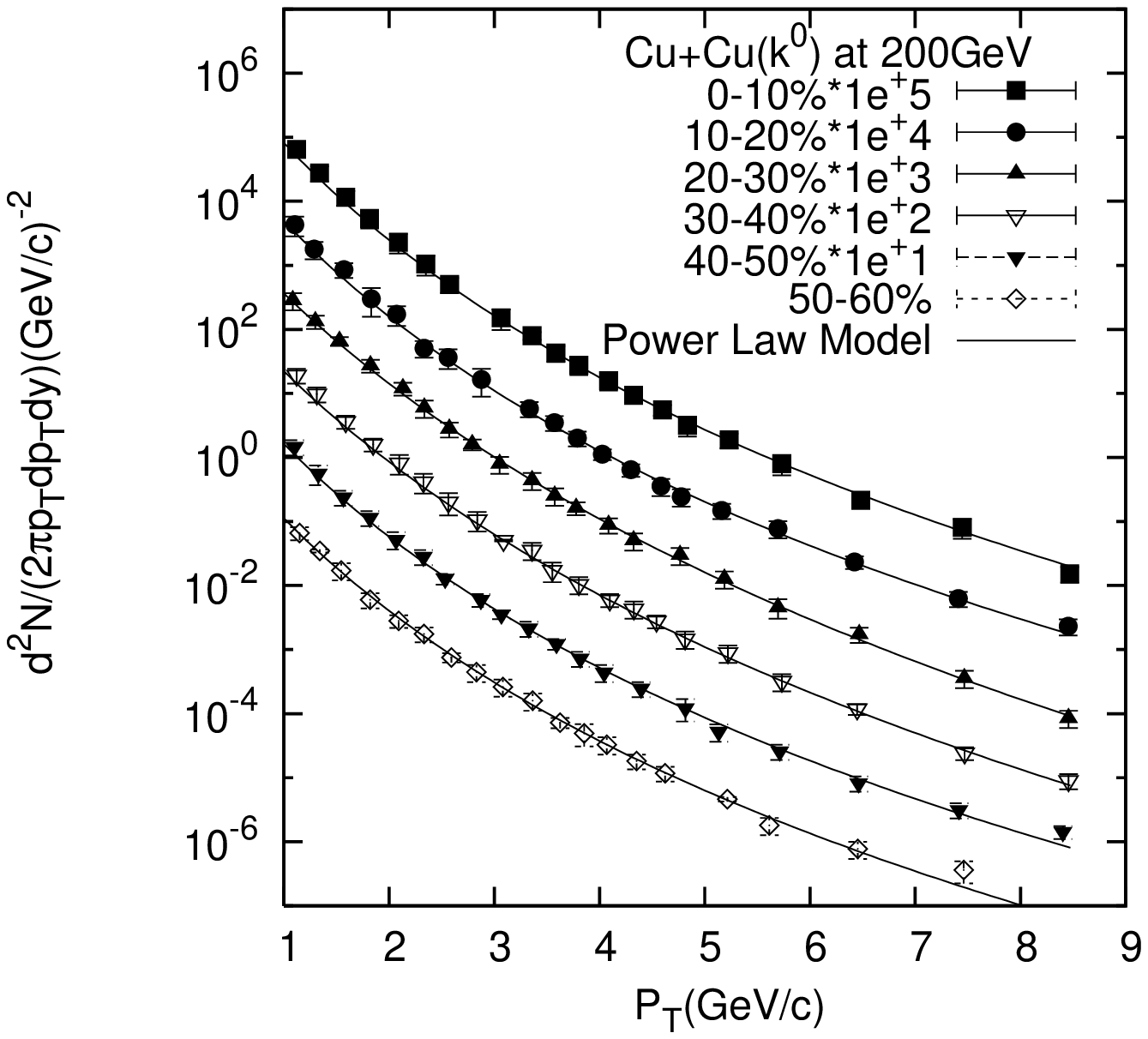}
\end{minipage}}%
\subfigure[]{
\begin{minipage}{.5\textwidth}
\centering
\includegraphics[width=2.5in]{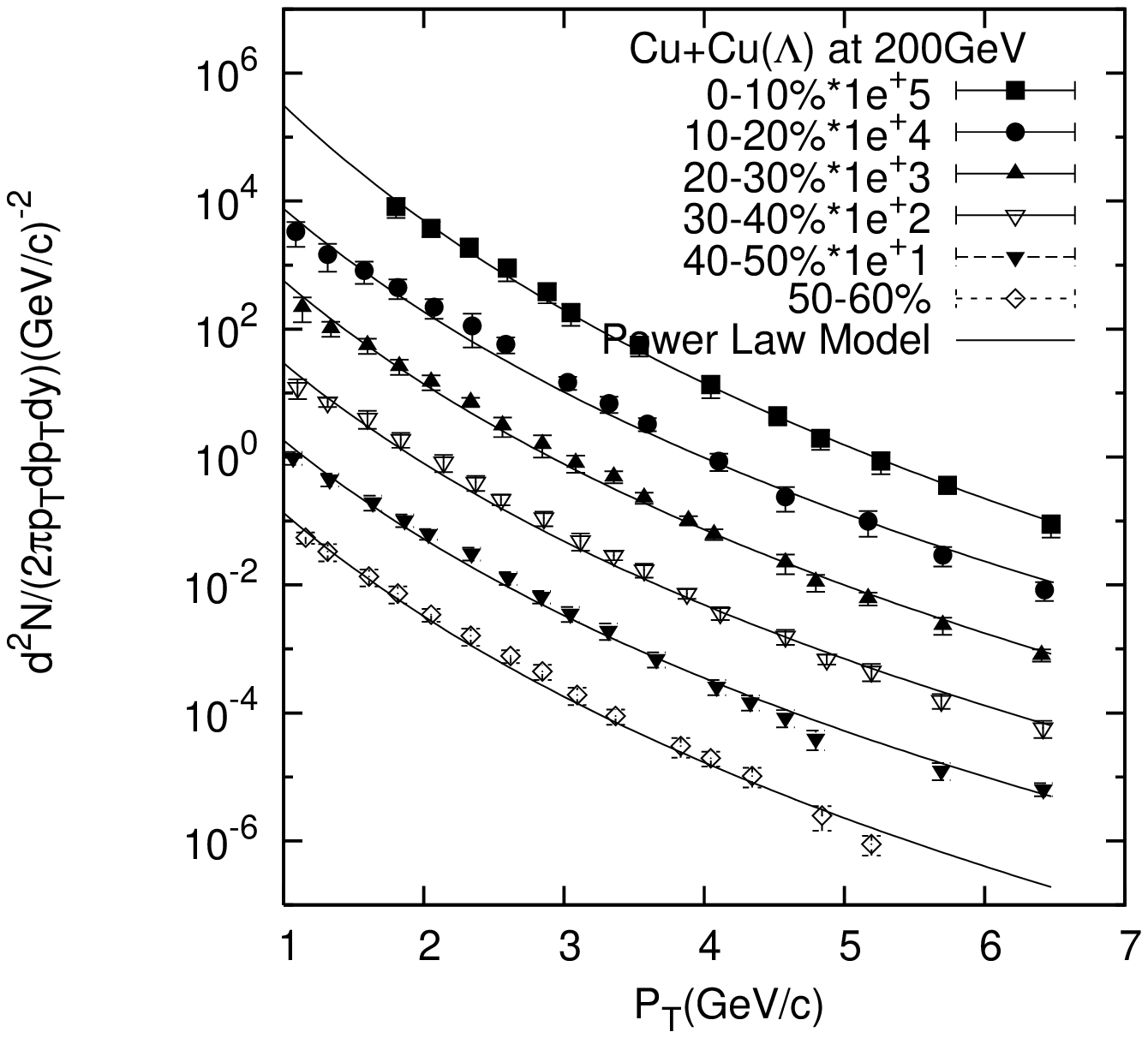}
\end{minipage}}%
\caption{Transverse momentum spectra for production of neutral keons$(k^0_s)$ and lamda($\Lambda$)
particles at different centrality in Au-Au collisions at $\surd{s}$=200GeV.
The experimental data are taken from Ref.\cite{Wang1} The
solid curves are fits for power-law model.}
\end{figure}

\begin{figure}
\subfigure[]{
\begin{minipage}{.5\textwidth}
  \centering
\includegraphics[width=2.5in]{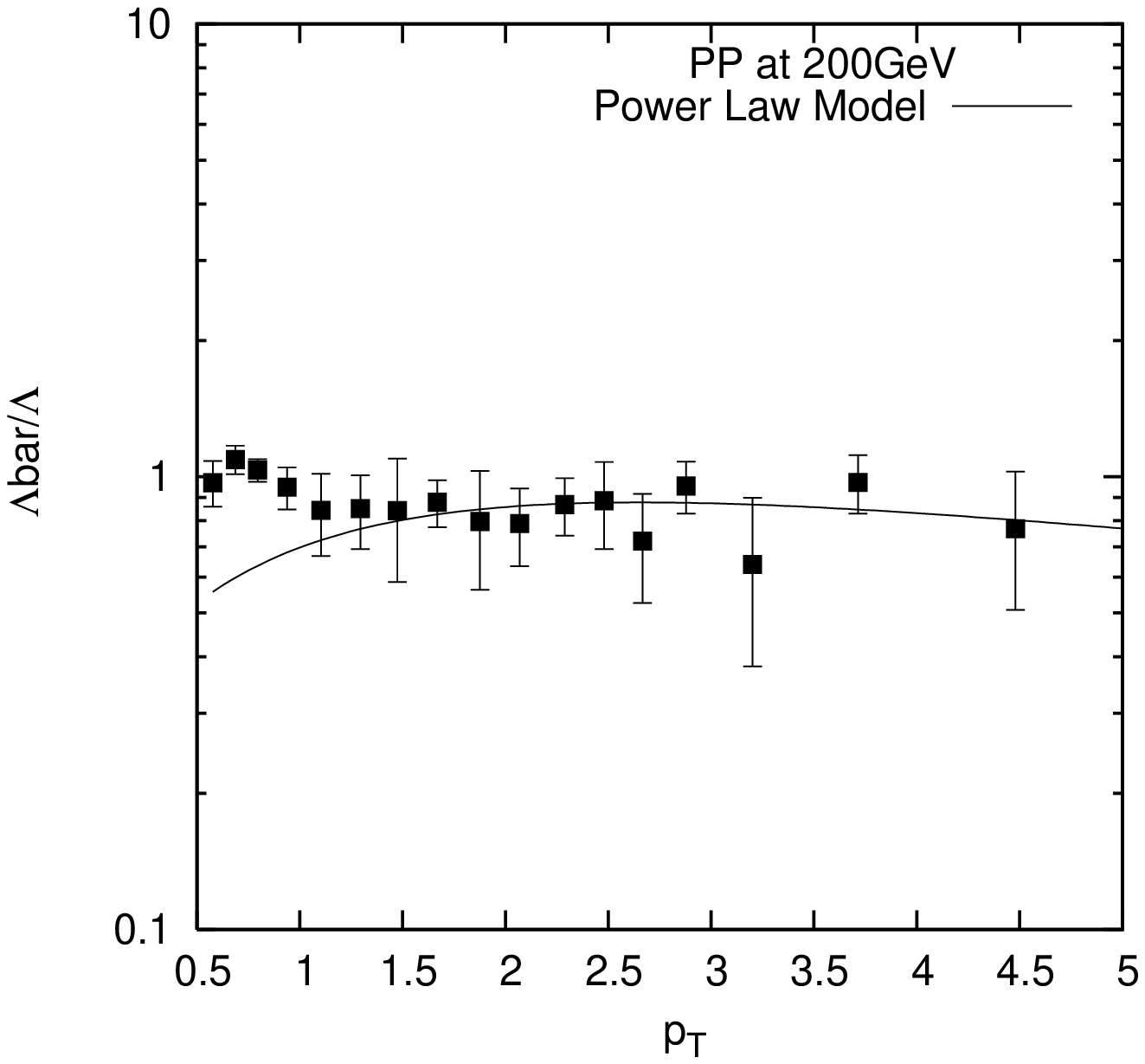}
\end{minipage}}%
\subfigure[]{
\begin{minipage}{.5\textwidth}
\centering
\includegraphics[width=2.5in]{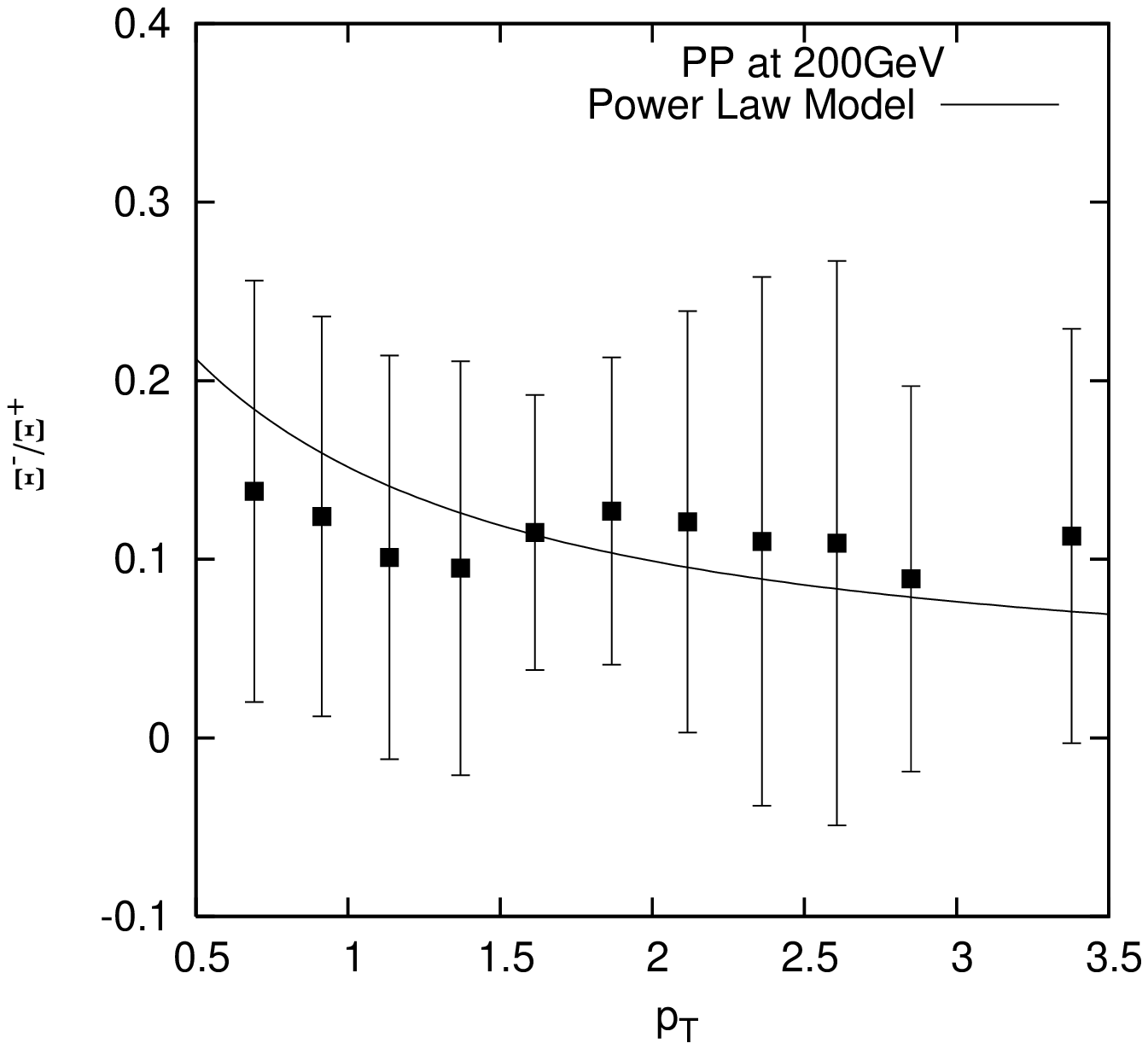}
\end{minipage}}%
\caption{Transverse momentum-dependence spectra of
$\Lambda$bar/$\Lambda$ and $\Xi^-$/$\Xi^+bar$ for pp collision at
$\surd{s}$=200GeVThe data type-points are taken from
Ref.\cite{Abelev1}. The solid curves or lines are drawn on the basis
of Power Law Model.}
\end{figure}

\begin{figure}
\subfigure[]{
\begin{minipage}{.5\textwidth}
  \centering
\includegraphics[width=2.5in]{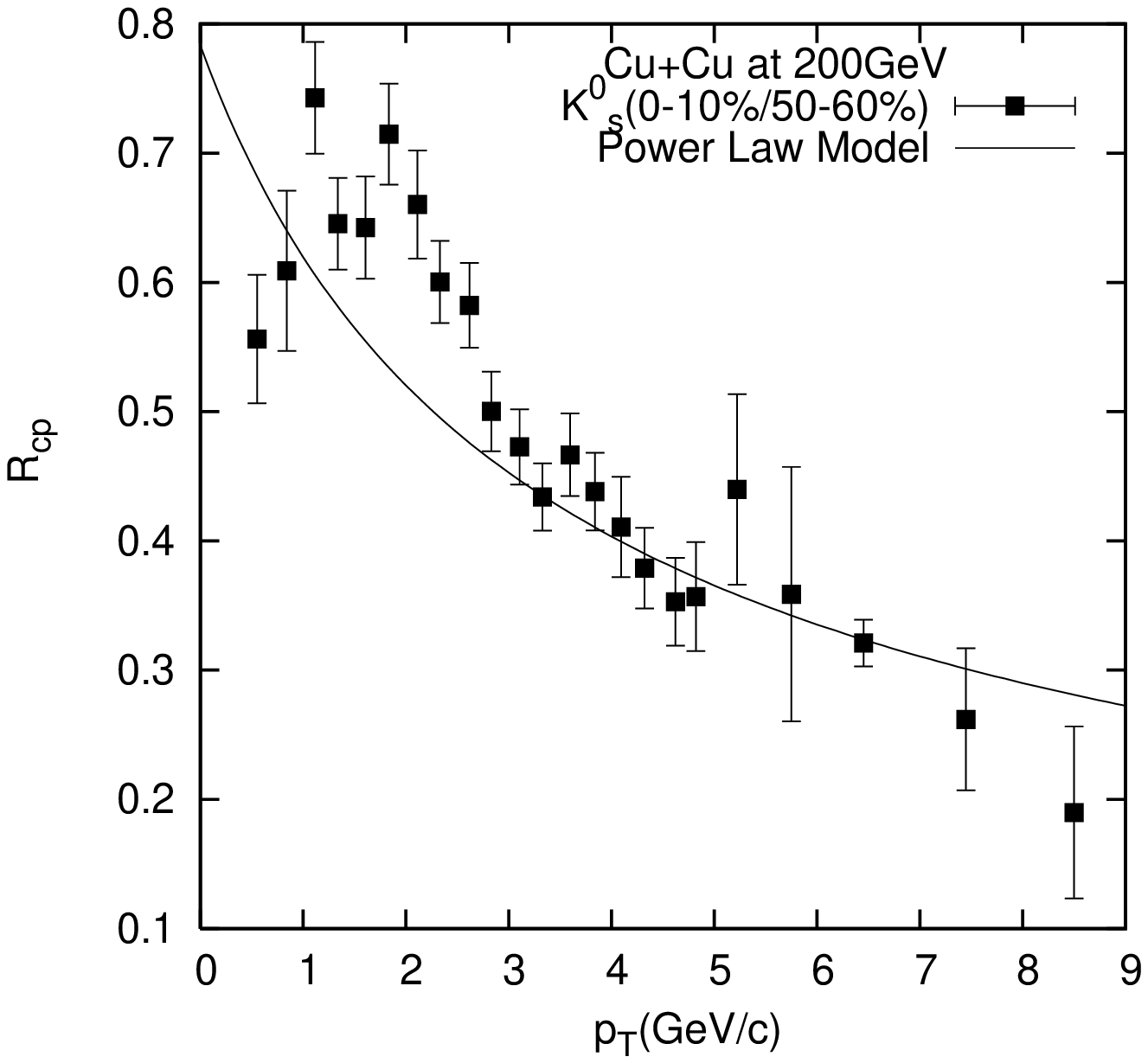}
\end{minipage}}%
\subfigure[]{
\begin{minipage}{.5\textwidth}
\centering
\includegraphics[width=2.5in]{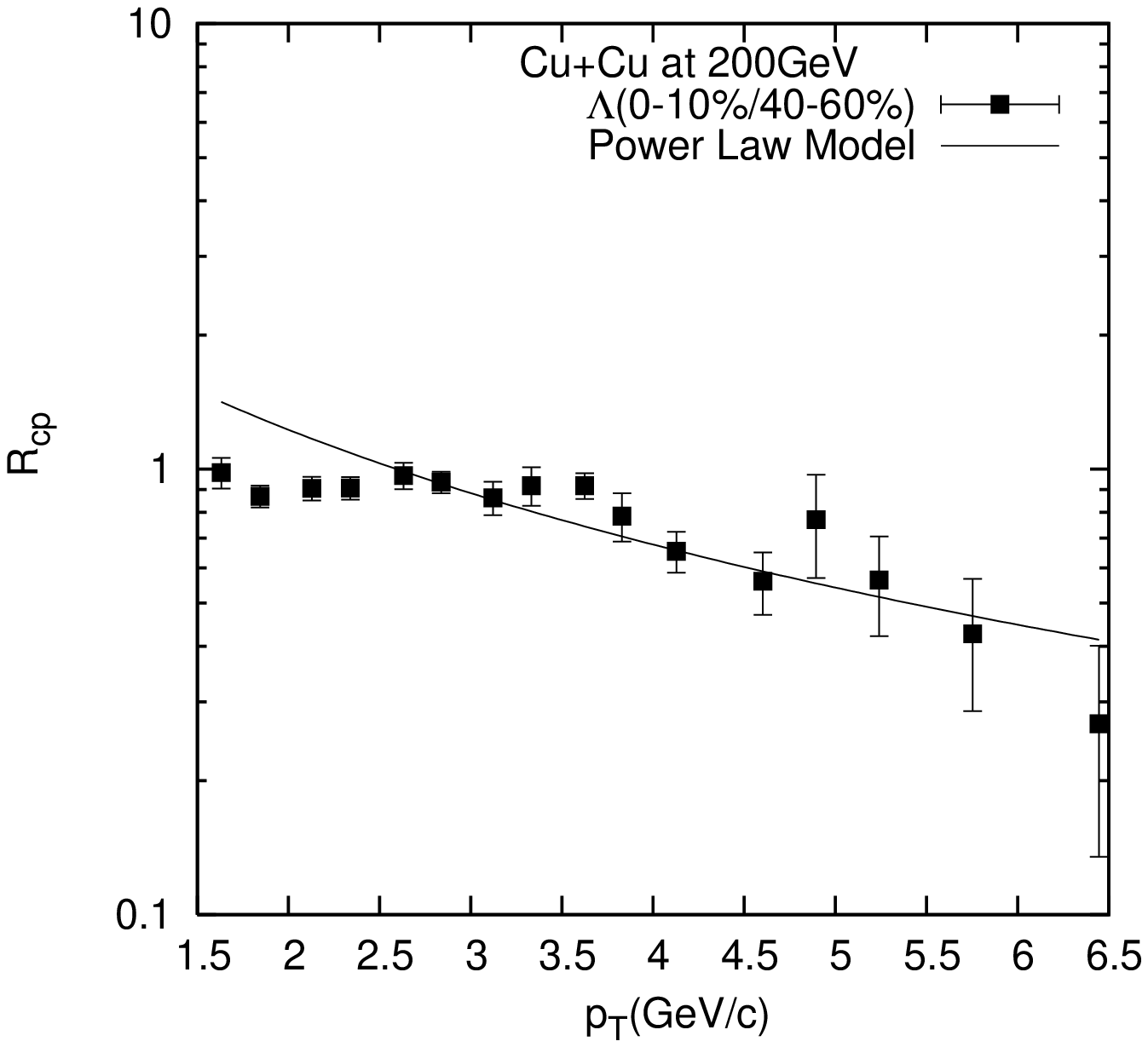}
\end{minipage}}%
\caption{Plots of the nuclear modification factor $(R_{cp})$ versus $p_T$
spectra for Cu-Cu collisions at $\surd{s}$=200GeV.
The experimental data are taken from Ref.\cite{Wang1} The
solid curves indicate the power-law-based description of the data.}
\end{figure}

\end{document}